\documentclass[superscriptaddress,secnumarabic,amssymb,amsmath,nobibnotes,aps,prd,nofootinbib,preprint]{revtex4}

\usepackage{graphicx}
\usepackage{float}
\usepackage{bm}
\usepackage{amsmath}
\usepackage{amsfonts}
\usepackage{amssymb}
\usepackage{epstopdf}
\usepackage{natbib}%
\newcommand{\bee}{\begin{equation}}
\newcommand{\eee}{\end{equation}}
\newcommand{\eaa}{\end{eqnarray}}
\newcommand{\baa}{\begin{eqnarray}}

\usepackage{color}

\begin{document}

\title{Note on an extended chiral bosons system contextualized in a modified gauge-unfixing formalism}
\author{Gabriella V. Ambrósio}
\email{gabriellambrosio@gmail.com}
\affiliation{Departamento de F\'isica, Universidade Federal de Juiz de Fora, Juiz de Fora – 36036-330, MG, Brazil}
\author{Cleber N. Costa}
\email{cleber.costa@ice.ufjf.br}
\affiliation{Departamento de F\'isica, Universidade Federal de Juiz de Fora, Juiz de Fora – 36036-330, MG, Brazil}
\author{Paulo R. F. Alves}
\email{paulo.alves@ice.ufjf.br}
\affiliation{Departamento de F\'isica, Universidade Federal de Juiz de Fora, Juiz de Fora – 36036-330, MG, Brazil}
\author{Jorge Ananias Neto}
\email{jorge@fisica.ufjf.br}
\affiliation{Departamento de F\'isica, Universidade Federal de Juiz de Fora, Juiz de Fora – 36036-330, MG, Brazil}
\author{Ronaldo Thibes}
\email{thibes@uesb.edu.br}
\affiliation{Universidade Estadual do Sudoeste da Bahia, DCEN,
Rodovia BR 415, Km 03, Itapetinga – 45700-000, BA, Brazil}

\begin{abstract}
We analyze the Hamiltonian structure of an extended chiral bosons theory in which the self-dual constraint is introduced via a control $\alpha$-parameter.
The system has two second-class constraints in the non-critical regime and an additional one in the critical regime.  
We use a modified gauge unfixing formalism to derive a first-class system, disclosing hidden symmetries. To this end, 
we choose one of the second-class constraints to build a corresponding gauge symmetry generator.
The worked out procedure converts second-class variables into first-class ones allowing the lifting of gauge symmetry. Any function of these GU variables will also be invariant.
We obtain the GU Hamiltonian and Lagrangian densities in a generalized context containing the Srivastava and Floreanini-Jackiw models 
as particular cases. Additionally, we observe that the resulting GU Lagrangian presents similarities to the Siegel invariant Lagrangian which is known to be suitable for describing chiral bosons theory with classical gauge invariance, however broken at quantum level. The final results signal a possible equivalence between our invariant Lagrangian obtained from the modified GU formalism and the Siegel invariant Lagrangian, with a distinct gauge symmetry.
\end{abstract}


\maketitle

Chiral bosons (CB) are basic ingredients in the description of two-dimensional
scalar field models \cite{Sevrin:2013nca, Mezincescu:2022hnb, epl3, Arvanitakis:2022bnr} with applications to modified gravity theories \cite{Giaccari:2008zx, Sen:2015nph, Merbis:2023uax}, noncommutativity \cite{epl3, Miao:2003ab, Das:2004vc} and string/brane theories \cite{Tseytlin:1990va, bar, Hull:2004in, Phonchantuek:2023iao}, besides providing building blocks for fundamental concepts in modern theoretical physics \cite{Abreu:2000sn, Upadhyay:2011ph, Shukla:2013baa}. Additionally, CB theories also present rich mathematical canonical structures {\it per si}, 
among which we underline the typical chiral constraint algebra that appears in the Floreanini-Jackiw (FJ) theory \cite{fjprl,bw,ea}. Interestingly enough, the
original FJ chiral constraint manifests itself in a single second-class form, with a non-trivial commutator along the real line. In higher dimensions, the chirality condition generalizes to chiral $p$-form theories with self-dual field strength, from which we mention the interesting Past-Sorokin-Tonin (PST) approach \cite{Pasti:1995ii, Pasti:1995tn, Pasti:1996vs}, recently refined by Mkrtchyan \cite{Mkrtchyan:2019opf},  where gauge-invariant self-dual models containing infinitely many auxiliary fields are related to equivalent ones with fewer finite auxiliary fields in non-polynomial form.  Neither infinite set of fields nor non-polynomiality being welcome properties, they seemed at first necessary for a consistent first-class theory.   Building on established string theory compactification ideas, over the past years, Ashoke Sen has shown that for specific space-time dimensions it is possible to avoid both situations obtaining a finite Lorentz covariant action containing an extra set of free unphysical fields \cite{Sen:2015nph, Sen:2019qit}.  Correspondingly, a partition function for Sen's action has been recently evaluated in \cite{Andriolo:2021gen, Lambert:2023qgs} providing further novel developments on the duality relations.  Following a different path, in the present note, we focus on the peculiarities of the two-dimensional case from the gauge-unfixing formalism viewpoint searching for hidden symmetries and clues to the quantization process.
A pioneering proposal for including the chirality condition in a two-dimensional scalar theory using an auxiliary Lagrange 
multiplier field was developed by Srivastava in \cite{sriva}. Unfortunately, Srivastava's model does not seem to be equivalent to the 
FJ one and does not even appear
to be unitary since its reduced Hamiltonian cannot be positive definite \cite{harada,ggr}. One way of circumventing this difficulty is by
introducing an additional open free parameter $\alpha$ along with the Lagrange multiplier field, leading to the Kim, Kim and Park (KKP) model \cite{kkp} which embodies the chiral constraint linearly into the Lagrangian.
On that case, depending on the value of the parameter $\alpha$, one can reproduce the FJ or the Srivastava models. That extended model amounts to
a second-class Dirac-Bergmann system with two second-class constraints.  On the other hand, the Siegel version for the CB \cite{sig} contains the chirality condition quadratically and enjoys gauge invariance, which is known to be related to first-class systems.  Inspired by those facts, in this current note, we propose to convert the second-class KKP model into first-class in order to analyze the extended CB system in the context of gauge theories.  For that purpose, we employ the modified gauge-unfixing (GU) formalism \cite{epl3, proto, epl4} to construct a broader gauge theory encompassing the extended CB as a particular gauge-fixing realization.  In addition to playing a key role in the description of all fundamental interactions, gauge theories are also technically significant as the quantization of second-class systems is substantially more complex than first-class ones \cite{mes}.  Furthermore, by exposing new symmetries, the conversion to first class often sheds light to open problems from different perspectives as it allows the application of a powerful machinery of functional quantization methods \cite{Mandal:2022xuw}. 
Along the last decades, two main second-to-first-class system conversion lines have evolved in the literature, leading to different formalisms. 
The first one comes from pioneering works by Wess, Zumino, Faddeev and Shatashvili \cite{Wess:1971yu, fasha} in which extra variables are introduced in phase space in order to convert second-class constraints into first-class and has led to the systematic Batalin-Fradkin-Tyutin (BFT) approach \cite{bfto1,bfto2} with corresponding subsequent modifications and improvements \cite{Banerjee:1994pp, Kim:2005mf, Pandey:2021myh}. Since then, there have been countless applications of the BFT formalism.   Closer to our present topic, we can particularly mention the work of Amorim and Barcelos-Neto in which the BFT formalism is directly applied to CB models \cite{Amorim:1994np, ab}. The second line, which we shall be following here, originated from works by Mitra, Rajamaran and Vytheeswaran \cite{mr, vyt1,vyt2,vyt3} and is based on more clean economical grounds as it does not rely on the introduction of new variables but rather views the second-class system as a gauge-fixed version of a more general first-class theory.  Bringing more understanding to the intricacies of singular Dirac-Bergmann systems,  this second approach has grown in depth along the literature finding many applications and has come to be known as the GU formalism.  As far as we know, apart from the recent work on noncommutativity \cite{epl3}, the GU formalism has never been applied to CB theories before.  Concerning the existing first and second class CB descriptions along with their corresponding mentioned concurring issues, the current GU approach to the quantization of an extended CB model encompassing the previous situations as particular limiting cases certainly brings new relevant features for CB theories.

In the original Vytheeswaran works \cite{vyt1,vyt2,vyt3}, only already existing phase space variables are used to convert second-class systems into first-class ones. No Wess-Zumino terms are introduced and a Lie projection operator is directly applied to the second-class Hamiltonian to reveal hidden symmetries.
The modified GU formalism \cite{epl3, proto, epl4} consists in redefining the second-class phase space variables as first-class ones.
It turns out that the construction of gauge-invariant variables from start leads to considerable simplifications when compared with the usual GU formalism.
Given a second-class system, we split the constraints into two families, the first of which shall provide the gauge symmetry generators while the second is actually discarded in a broader gauge-invariant theory context in which they play a role of possible gauge choices among others.
Consider, for example, a second-class phase-space function $T(A_{\mu},\pi_{\mu})$, with the index $\mu$ running through all phase space variables.  The strategy is to write a corresponding first-class version $\tilde{T}({A}_{\mu},{\pi}_{\mu})$ obtained from the second-class one $T$ with the same functional form as
\begin{equation}
    \tilde{T}({A}_{\mu},{\pi}_{\mu}) \equiv {T}(\tilde{A}_{\mu},\tilde{\pi}_{\mu})
    \,,
\end{equation}
by redefining the original phase space variables
\begin{equation}
    A_\mu \longrightarrow \tilde{A}_\mu ({A}_{\mu},{\pi}_{\mu}) \,,
\end{equation}
\begin{equation}
    \pi_\mu \longrightarrow \tilde{\pi}_\mu ({A}_{\mu},{\pi}_{\mu}) \,,
\end{equation}
such that
\begin{equation}
  \delta \tilde{A}_\mu = \alpha \left \{ \tilde{A}_\mu , \psi \right \}=0 \,, \label{3.8}
\end{equation}
\begin{equation}
  \delta \tilde{\pi}_\mu = \alpha \left \{ \tilde{\pi}_\mu , \psi \right \}=0 \,, 
\end{equation}
where $\alpha$ are the infinitesimal gauge parameters and $\psi$ the gauge symmetry generators built from the second-class constraints. The new deformed variables $\tilde{A}_\mu$, $\tilde{\pi}_\mu$ are known as GU variables. It is clear that all functions of the GU variables, in particular $\tilde{T}$, will be gauge 
invariant since
\begin{align}
\left \{ \tilde{T}, \psi \right \}= \left \{ \tilde{A}, \psi \right \} \frac{\partial  {T} }{\partial \tilde{A}} + \frac{\partial {T}}{\partial \tilde{\pi }}\left \{ \tilde{\pi }, \psi \right \} = 0 \,.
\end{align}
Consequently, we can obtain a gauge invariant function from the replacement
\begin{align}
\label{repla}
T(A_{\mu}, \pi_{\mu}) \rightarrow T(\tilde{A_{\mu}}, \tilde{ \pi_{\mu}})=\tilde{T}({A_{\mu}},{ \pi_{\mu}}) \,.
\end{align}
For the sake of argument clearness, assume for the moment that the system has only two second-class constraints $Q_1$ and $Q_2$. The phase space gauge invariant GU variables, collectively denoted by $\tilde{\Lambda}\equiv (\tilde{A}_\mu,\tilde{\pi}_\mu)$, can be constructed as a power series in the discarded second-class constraint $Q_{2}$ as
\begin{align}
\tilde{\Lambda}(x)=\Lambda(x)+\int dyb_{1}(x,y)Q_{2}(y)+\iint dydz b_{2}(x,y,z)Q_{2}(y)Q_{2}(z)+... \,. \label{3.11}
\end{align}
This assures that we recover the original second-class system back when $Q_2=0$ as the GU variables satisfy the boundary condition
\begin{align}
    {\tilde{\Lambda}}_{\big| Q_{2}=0}=\Lambda
\end{align}
on the constraint surface $Q_{2}=0$.
The unknown coefficient functions $b_{n}$ present in Eq. (\ref{3.11}) are then determined by the GU gauge invariant condition
\begin{equation}\label{GUIC}
\delta\tilde{\Lambda}= \alpha \left \{ \tilde{\Lambda} , Q_1 \right \}=0\,.
\end{equation}
The general equation for $b_{n}$ obtained from (\ref{GUIC}) reads
\begin{align}
\delta \tilde{\Lambda}(x)=\delta \Lambda(x)+\delta \int dyb_{1}(x,y)Q_{2}(y)+ \delta \iint dydz b_{2}(x,y,z)Q_{2}(y)Q_{2}(z)+...=0 \,, \label{3.18}
\end{align}
with the infinitesimal transformations being given by 
\begin{align}
 & \delta \Lambda(x)=\int dy \, \alpha(y)\left \{ \Lambda(x), \psi(y) \right \}, \\
 &\delta b_{1}(x)=\int dy \, \alpha(y)\left \{ b_{1}(x), \psi(y) \right \},\\
&\delta Q_{2}(x)=\int dy \, \alpha(y)\left \{ Q_{2}(x), \psi(y) \right \} \,.
\end{align}
So, for the first order correction term $(n = 1)$, we have from Eq.~\eqref{3.18} 
\begin{align}
\label{b1s}
    \delta \Lambda(x)+\int dyb_{1}(x,y)\delta Q_{2}(y)= 0 \,.
\end{align}
From Eq. ({\ref{b1s})} we can determine $b_1$.
For the second order correction term $(n = 2)$, we have
\begin{eqnarray}
\label{b2s}
    \int dy\delta b_{1}(x,y)Q_{2}(y)+2\iint dydz b_{2} (x,y,z)\delta Q_{2}(y)Q_{2}(z)= 0 \,.
\end{eqnarray}
Then, from Eq. ({\ref{b2s})} we can determine $b_2$ and the same procedure described so far is used to 
determine all the remaining $b_n$ coefficient functions.
Therefore, from the GU variables power series defined in Eq.~(\ref{3.11}), we can derive a corresponding gauge invariant theory.

We proceed now to our main course in which we consider an extended CB model described by a Lagrangian density obtained from the sum of a free scalar field theory with the modified linear self-dual constraint \cite{kkp} 
\begin{equation}
\label{ls}
    \mathcal{L}_\alpha=\frac{1}{2}(\dot{\phi}^{2} -{\phi}'^{2}) +\lambda(\dot{\phi}-{\phi}')+\frac{1}{2}\alpha\lambda^{2} \,.
\end{equation}
The overdot and prime mean differentiation with respect to time $t$ and space $x$, respectively,
$\lambda$ stands for an auxiliary field working as a Lagrange multiplier and $\alpha$ is a convenient regularizing parameter which may, as we shall soon see, succeed in solving the difficulties present in Srivastava's model \cite{sriva}. The point is that (\ref{ls}) smoothly interpolates among some interesting relevant values of $\alpha$.  For $\alpha\rightarrow0$, it describes Srivastava's chiral boson with the chirality condition arising naturally as the equation of motion for $\lambda$ while, as we shall see throughout the article, for $\alpha\rightarrow1$ we have a FJ chiral boson whilst in the improper limit
$\alpha\rightarrow \infty$ the chirality condition is suppressed leading to a usual free scalar field.

From the Lagrangian density (\ref{ls}), we get the momenta field variables
\begin{align}
\pi_{\lambda} = \frac{\partial \mathcal{L}_\alpha}{\partial \dot{\lambda}} & = 0 \label{a} \,,\\
\pi_{\phi} = \frac{\partial \mathcal{L}_\alpha}{\partial \dot{\phi}} & = \dot{\phi} + \lambda \,,
\end{align}
which lead to the primary constraint
\begin{eqnarray}
\label{omega1}
\Omega_{1} \equiv \pi_{\lambda} \approx  0 \,. \label{b} 
\end{eqnarray}
Performing a Legendre transformation
\begin{eqnarray}
\mathcal{H}_{\alpha} = \pi_{\phi}\dot{\phi} - \mathcal{L}_\alpha   \,,
\end{eqnarray}
we obtain the canonical Hamiltonian
\begin{align}
\label{ch}
\mathcal{H}_{\alpha} = \frac{1}{2} \pi^{2}_{\phi} + \frac{1}{2} {\phi}'^{2} - \lambda(\pi_{\phi} -{\phi}')-\frac{1}{2}(\alpha-1)\lambda^{2} \,.
\end{align}
Following the usual additional steps of the Dirac-Bergmann algorithm
\cite{Dirac:1950pj, Anderson:1951ta}, the dynamical stability of $\Omega_{1}$ generates a further secondary constraint given by
\begin{eqnarray}
\label{omega2}
\Omega_{2} \equiv \pi_{\phi} -{\phi}' + (\alpha-1)\lambda  \approx 0 \, . \label{c} 
\end{eqnarray}
Since $\lambda$ has a non-null Poisson bracket with $\pi_\lambda$, the Dirac-Bergmann iterative process comes to an end with $\Omega_{1}$ and $\Omega_{2}$ representing the complete set of constraints for the model. The nature of the system is revealed from the corresponding Poisson bracket relations among the constraint functions
\begin{align}
\label{o22}
\left\{\Omega_{2}(x), \Omega_{2}(y) \right\} &= - 2 \delta'(x-y) \,, \\
\label{alfa}
\left\{\Omega_{1}(x), \Omega_{2}(y) \right\} &= -(\alpha - 1)\delta(x-y) \,.
\end{align}
For $\alpha \neq 1$, the constraints $\Omega_{1}$ and $\Omega_{2}$ satisfy a second-class algebra. From the previous brackets,
defining $C_{ij}(x,y)\equiv\left \{ \Omega_{i}(x), \Omega_{j}(y) \right \}$, with $i,j=1,2$, and $C^{-1}_{ij}(x,y)$ by
\begin{equation}
    \int dz \,
    \left\{
    \,\Omega_i(x)\,,\,\Omega_k(z)\,
    \right\}
    C_{kj}^{-1}(z,y) = 
    \int dz \,
    C_{ik}^{-1}(x,z)
    \left\{
    \,\Omega_k(z)\,,\,\Omega_j(y)\,
    \right\}
     = \delta_{ij}\delta(x-y)
    \,,
\end{equation}
we obtain the Dirac matrix $C_{ij}(x,y)$
and its corresponding inverse $C^{-1}_{ij}(x,y)$ as
\begin{align}
C_{ij}(x,y) = \begin{bmatrix}
0 &  1- \alpha \\ 
 \alpha - 1 & -2\partial_{x}
\end{bmatrix}\delta(x-y)
\end{align}
and
\begin{align}\label{IDM}
C^{-1}_{ij}(x,y) =\frac{1}{\alpha - 1}\begin{bmatrix}
-2\partial_{x} &  1 \\ 
  - 1 & 0
\end{bmatrix} \delta(x-y)\,,
\end{align}
which are needed to compute de Dirac brackets (DB) algebra. In fact, by inserting the inverse Dirac matrix (\ref{IDM}) into the general definition
\begin{align}\label{DB}
    \{F(x),G(y)\}_D = \{F(x),G(y)\} 
    -\int dz d\bar{z} \, \{F(x),\Omega_i(z)\}C^{-1}_{ij}(z,\bar{z})\{\Omega_j(\bar{z}),G(y)\}
    \,,
\end{align}
the fundamental DB's among the phase space variables can be computed as
\begin{align}
&\left \{ \phi(x), \pi_{\phi}(y) \right \}_{D}=\delta(x-y) \,,\\
&\left \{ \phi(x), \phi(y) \right \}_{D}=\left \{ \pi_{\phi}(x), \pi_{\phi}(y) \right \}_{D}=0 \,,\\
&\left \{ \phi(x), \lambda(y) \right \}_{D}=-\frac{1}{\alpha -1}\delta(x-y) \,.
\end{align}
At this moment, one can proceed with the canonical quantization program by ensuring that the corresponding operators adhere to commutation relations as established by the
DB algebra obtained above. In addition, other functional quantization framework lines can be used. A key point in the GU formalism amounts to reproducing the precise above DB structure by means of ordinary PBs among the GU variables.  We shall come to that fine result in a moment.

In order to apply the GU technique to the current model, it is convenient to rescale the primary constraint $\Omega_1=\pi_{\lambda}$ to construct a proper gauge symmetry generator
\begin{eqnarray}
\label{g1s}
\psi \equiv -\frac{\Omega_{1}}{(\alpha -1)}= -\frac{\pi_{\lambda}}{(\alpha -1)} \,,
\end{eqnarray}
so that $\left \{ \psi(x), \Omega_{2}(y) \right \}=\delta(x-y)$. The infinitesimal gauge transformations generated by $\psi$ are
\begin{align}
\label{gt1}
\delta \phi(x)&=\int dy \varepsilon (y)\left \{ \phi(x), \psi(y) \right \}=0 \,, \\
\label{gt2}
\delta \pi_{\phi}(x)&= \int dy \varepsilon (y)\left \{ \pi_{\phi}(x), \psi(y) \right \}=0 \,, \\
\label{gt3}
\delta \lambda(x)&= \int dy \varepsilon (y)\left \{ \lambda(x), \psi(y) \right \}=-\frac{1}{(\alpha - 1)}\varepsilon (x) \,, \\
\label{f}
\delta \Omega_{2}(x)&=\int dy \varepsilon (y)\left \{ \Omega_{2}(x), \psi(y) \right \} 
=-\varepsilon (x) \,.
\end{align}
Thus, we can see that precisely $\lambda$ and $\Omega_2$ are not invariant.  The corresponding GU gauge invariant variable $\Tilde{\lambda}$ can be constructed in terms of a power series in $\Omega_{2}$
\begin{eqnarray}
\label{a0t}
\tilde{\lambda}(x)= \lambda(x)+\int dy b_{1}(x,y)\Omega_{2}(y)+\iint dy dz b_{2}(x,y,z)\Omega_{2}(y)\Omega_{2}(z)+... \,.
\end{eqnarray}
Applying the GU invariant condition $\delta \tilde{\lambda}=0$, we can compute all the coefficient functions  $b_{n}$ corresponding to the correction terms in the series (\ref{a0t}). For the $\Omega_{2}$ linear order correction, we obtain
\begin{eqnarray}
\label{da0}
\delta\lambda(x)+\int dy  b_{1}(x,y)\delta \Omega_{2}(y)=0 \,, \label{g}
\end{eqnarray}
from which, by using Eqs. \eqref{gt3} and \eqref{f}, follows
\begin{eqnarray}
b_{1}(x,y)= -\frac{1}{(\alpha  -1)}\delta(x-y) \,. \label{h}
\end{eqnarray}
For the quadratic term we have $b_2=0$ and it is clear that all the remaining correction coefficient functions $b_{n}$ are null for $n\geq 2$. Consequently, considering this fact and plugging Eq.~\eqref{h} into (\ref{a0t}), we can derive the GU variable $\tilde{\lambda}$ as
\begin{eqnarray}
\label{a0t1}
\tilde{\lambda}(x)=\lambda(x)-\frac{1}{(\alpha -1)}\Omega_{2}(x) \nonumber\\ = - \frac{1}{\alpha-1} ( \pi_\phi - \phi' ) \,.
\end{eqnarray}
Therefore, we have achieved our first task obtaining the deformed GU variables 
\begin{align}
\label{fit}
&\tilde{\phi}=\phi\,,\\
\label{pit}
&\tilde{\pi}_{\phi}=\pi_{\phi}\,,\\
\label{lt}
&\tilde{\lambda}= - \frac{1}{\alpha-1} ( \pi_\phi - \phi' ) \,.
\end{align}

The canonical Poisson bracket relations among the GU variables are given by
\begin{align}
&\left \{ \tilde{\phi}(x), \tilde{\pi}_{\phi}(y) \right \}=\delta(x-y)=\left \{ \phi(x), \pi_{\phi}(y) \right \}_{D} \,,\\
&\left \{ \tilde{\phi}(x), \tilde{\phi}(y) \right \}=\left \{ \tilde{\pi}_{\phi}(x), \tilde{\pi}_{\phi}(y) \right \}=0=\left \{ \phi(x), \phi(y) \right \}_{D}=\left \{ \pi_{\phi}(x), \pi_{\phi}(y) \right \}_{D} \,,\\
&\left \{ \tilde{\phi}(x), \tilde{\lambda}(y) \right \}=-\frac{1}{\alpha -1}\delta(x-y) = \left \{ \phi(x), \lambda(y) \right \}_{D} \,.
\end{align}
As we can see, the above Poisson brackets structure satisfied by the GU variables completely agree with the previous results obtained in terms of DB's calculated from the definition (\ref{DB}) using the original phase space variables. This fine result, as anticipated in \cite{db}, provides an important calculational shortcut and confirms the consistency of the improved GU formalism.

Replacing the second-class variables $\phi, \pi_\phi$ and $\lambda$ by the GU variables $\tilde{\phi}, \tilde{\pi}_\phi$ and $\tilde{\lambda}$,
given by Eqs. (\ref{fit}), (\ref{pit}) and (\ref{lt}) respectively, 
in the canonical Hamiltonian, Eq. (\ref{ch}), we can derive the gauge invariant Hamiltonian density
\begin{align}
\tilde{\cal H}_\alpha &= \frac{1}{2} \pi_\phi^2 + \frac{1}{2} \phi'\,^2 + \frac{1}{2} (\alpha - 1)^{-1} (\pi_\phi - \phi')^2 \nonumber \\
& = \frac{1}{2} \frac{\alpha}{(\alpha -1)} \left ( \pi_{\phi} - \phi' \right )^{2} + \pi_{\phi}\phi' . \label{i}
\end{align}
The GU Hamiltonian obtained above, Eq. (\ref{i}), coincides with the second-class reduced Hamiltonian obtained in \cite{kkp}
and, as we can observe, it is not possible to guarantee that $\tilde{\cal H}_\alpha$
will be positive for $\alpha < 1$. This result is in line with those
previously discussed in \cite{harada,ggr,kkp}, regarding the fact that it is not possible to secure a positive definite Hamiltonian for Srivastava's model ($\alpha = 0$).

In order to obtain a well-behaving limit $\alpha\rightarrow 1$ for the gauge symmetry generator, we have to consider the vanishing of the numerator, $\pi_\lambda$,
in the corresponding term in Eq. (\ref{g1s}), which is precisely the primary constraint $\Omega_1 \equiv \pi_\lambda \approx 0$.
For the GU Hamiltonian, we have to consider the vanishing of the numerator, 
$ ( \pi_{\phi} -\phi')^2 $, in the corresponding term in Eq. (\ref{i}), to obtain a well-behaving limit $\alpha\rightarrow1$, which results the secondary constraint, Eq. (\ref{omega2}), 
\begin{eqnarray}
\label{qc1}
 \Omega_{2 (\alpha\rightarrow 1)}  \equiv \pi_{\phi} -\phi' \approx 0 \,.
\end{eqnarray}
Actually, condition (\ref{qc1}) above is already necessary for a consistent behavior of the GU variable $\tilde{\lambda}$ in that critical limit, c.f. Eq. (\ref{lt}).
The time evolution of (\ref{qc1}) produces one additional constraint
\begin{eqnarray}
\label{qce}
 \Omega_3  \equiv \lambda \approx 0 \,.
\end{eqnarray}
Further time evolution of $\Omega_3$ results in no additional constraints.
In our GU model, we have the fields $\phi, \pi_\phi, \lambda, \pi_\lambda$, three second-class constraints, which are Eqs. (\ref{omega1}), (\ref{qc1}) and (\ref{qce}), 
resulting in the number of 
degrees of freedom (DOF) being equal to 1 which confirms that, for $\alpha = 1$, our theory is equivalent to the FJ model. 
This same result was already obtained by KKP \cite{kkp} in a second-class extended CB system.

If we consider $\alpha = 0$ in Eq. (\ref{i}), we obtain the particular Hamiltonian
\begin{eqnarray}
\label{i0}
\tilde{\mathcal{H}}_{0} = \pi_{\phi}\phi' \,.
\end{eqnarray}
In this case, we have the same four fields, namely $\phi, \pi_\phi, \lambda, \pi_\lambda$, and one first-class constraint, $\pi_\lambda$,
resulting in DOF=2 and confirming that, for $\alpha = 0$, our model is equivalent to Srivastava's.
Still in the $\alpha=0$ case, by using the Hamilton equations of motion and performing an inverse Legendre transformation $\tilde{\mathcal{L}}_0 =   \pi_{\phi}\dot{\phi} - \tilde{\mathcal{H}}_{0}  \,$  in Eq. (\ref{i0}), we obtain the null result
$
 \tilde{\mathcal{L}}_0 = 0
$,
plainly agreeing with reference \cite{ggr}.

Back to the general case, calculating the Hamilton equation of motion for an arbitrary value of the parameter $\alpha$
\begin{eqnarray}
\label{hmi}
\frac{\delta \tilde{\mathcal{H}}_\alpha}{\delta \pi_{\phi}}= \frac{\alpha}{(\alpha -1)} \left ( \pi_{\phi} - \phi' \right )+\phi' 
= \dot{\phi} \,,
\end{eqnarray}
and using an inverse Legendre transformation, we can derive the GU Lagrangian 
\begin{align}
\label{li}
\tilde{\mathcal{L}}_\alpha &= \frac{1}{2\alpha} \left[ (\alpha - 1)  \dot{\phi} \dot{\phi} + 2 \dot{\phi} \phi' - (1 + \alpha) \phi'\phi'
\right] \nonumber \\ \nonumber\\
&= \frac{1}{2} \left[ ( \dot{\phi}^2 - {\phi'}^2 )  - \frac{1}{\alpha} ( \dot{\phi} - \phi' )^2 \right]\,.
\end{align}
Here it can be mentioned that, although the initial Lagrangian, Eq. (\ref{ls}), depends on the auxiliary field $\lambda$, the invariant Lagrangian, Eq. (\ref{li}), does not. Moreover, 
the GU invariant Lagrangian obtained above presents a quadratic term on the chirality condition and, consequently, the GU invariant Lagrangian 
has the same functional form as the invariant Lagrangian proposed by Siegel to describe chiral 
bosons \cite{sig,is}. However, the gauge transformations for the GU variables, c.f. Eqs. (\ref{gt1}), (\ref{gt2}) and (\ref{gt3}), are clearly different from Siegel's, which is known to be anomalous due to its gauge invariance breaking at quantum level. With the choice of $\Omega_1$  for constructing the gauge-symmetry generator in (\ref{g1s}), we see that while $\phi$ and $\pi_\phi$ do not transform, we have freedom to neatly gauge away the Lagrange multiplier from the theory.  The quest of whether such process can be done at quantum level in a non-anomalous way is currently under investigation.

Another way to understand the GU Lagrangian follows from the procedure where
the original second-class variables in the initial Lagrangian, Eq. (\ref{ls}), which are $\phi$ and $\lambda$, are replaced directly with the GU variables which are $\tilde{\phi}$ and $\tilde{\lambda}$,
given by Eqs. (\ref{fit}) and (\ref{lt}) respectively. Then, using the Hamiltonian equation of motion, Eq. (\ref{hmi}), 
with the aim of eliminating the momentum variable $\pi_\phi$, in which the GU variable  $  \tilde{\lambda}$, 
as we can see, is written in terms of it,
we again obtain the GU Lagrangian, Eq. (\ref{li}). 
Thus, we have established a remarkable condition on the invariant Lagrangian within the extended chiral boson model and modified GU formalism. Specifically, when employing the transformation-invariant variable $\phi$ and the transformation-exhibiting variable  
$\lambda$, the sole Lagrangian that preserves invariance seems to be uniquely determined by Eq. (\ref{li}).

For the limit $\alpha\rightarrow0$, if we consider the general GU Lagrangian (\ref{li}), we have 
\begin{align}
\label{li0}
\tilde{\mathcal{L}}_{\alpha\rightarrow0} = \lim_{\alpha \rightarrow 0} \; \frac{1}{2} \left[  ( \dot{\phi} - \phi')( \dot{\phi} + \phi') -
\frac{1}{\alpha} ( \dot{\phi} - \phi' )^2 \right] \,.
\end{align}
From Eq. (\ref{li0}), we can see that we have to impose the chirality condition $\dot{\phi} = \phi'$ in order 
to have the second term in the GU Lagrangian well-behaved. This chirality condition just corresponds to Srivastava's model \cite{sriva}.
In this way, Eq. (\ref{li0}) reduces to
\begin{align}
\label{li02}
\tilde{\mathcal{L}}_{\alpha\rightarrow0} = - \lim_{\alpha \rightarrow 0} \; \frac{1}{2\alpha} ( \dot{\phi} - \phi' )^2 \,.
\end{align}
Applying L'Hôpital's rule to calculate the limit in (\ref{li02}), we have $  \tilde{\mathcal{L}}_{\alpha\rightarrow0} = 0 $.
This result agrees with the one previously obtained performing a direct Legendre transformation to the $\alpha=0$ invariant Hamiltonian, Eq. (\ref{i0}).

If we set $\alpha = 1$, then the general GU Lagrangian, Eq. (\ref{li}), results in
\begin{equation}
\tilde{\mathcal{L}}_{1} =\dot{\phi}\phi'-\phi'^{2} \,,
\end{equation}
which is precisely the Lagrangian density for the Floreanini-Jackiw model \cite{fjprl}. For the case $\alpha = \pm \infty$ in (\ref{li}), we retrieve
a free scalar theory or the left-right chiral boson theory.
It is worth mentioning here that the secondary constraint, Eq. (\ref{omega2}), due to the noncommutative property, Eq. (\ref{o22}), is a single 
second-class constraint. Therefore, 
it is not possible to use it to build another gauge symmetry generator in the GU formalism, as it was done for the primary constraint $\Omega_1$, Eq. (\ref{omega1}).

In this note, we have converted an extended CB theory with a modified linear constraint into a first-class
system. We have used a modified GU formalism where the phase space variables are redefined as first-class variables
without introducing any Wess-Zumino terms. Consequently, any function of these first-class variables is gauge invariant.
In general, the procedure of obtaining GU variables simplifies the derivation of invariant quantities when compared with the usual 
GU formalism.
The general GU invariant Hamiltonian and Lagrangian densities, respectively Eqs. (\ref{i}) and (\ref{li}), were obtained in terms of the parameter $\alpha$. For $\alpha$ equal to zero, we have 
the Srivastava model. For $\alpha$ equal to one, we have the Floreanini-Jackiw model. Finally we can mention that once the GU 
variables are defined then they will be the ones that will reveal the properties of chiral bosons theory in a gauge invariant form. 
As a natural sequel of the present work, we are currently investigating whether the GU procedure applied to the extended chiral boson system leads to a theory that can be free or not of quantum anomalies.

We would like to thank the anonymous Referee for valuable suggestions.
We would also like to acknowledge CAPES (Coordenação de Aperfeiçoamento de Peso de Nível Superior) and FAPEMIG (Fundação de Amparo à Pesquisa do Estado de Minas Gerais) for financial support. Jorge Ananias Neto thanks CNPq (Conselho Nacional de Desenvolvimento Cient\'ifico e Tecnol\'ogico), Brazilian scientific support federal agency, for partial financial support, CNPq-PQ, Grant number 307153/2020-7.

\end{document}